\documentclass[runningheads]{llncs}
\usepackage[T1]{fontenc}

\usepackage{graphicx}
\usepackage{comment}
\usepackage{url}
\usepackage{hyperref}
\begin{document}
\title{Analyzing Middle School Students' Dialogue and Behaviors during Collaborative AI Chatbot Development Using Ordered Network Analysis}
%
%
\titlerunning{Analyzing Collaborative AI Chatbot Development}

%
\author{
Shan Zhang\inst{1}\orcidID{0009-0003-3532-0661} \and
Andres Felipe Zambrano\inst{2}\orcidID{0000-0003-0692-1209} \and
Xiaoyi Tian\inst{3}\orcidID{0000-0002-5045-0136} \and
Yukyeong Song\inst{4}\orcidID{0000-0002-4084-2734} \and
Anthony F. Botelho\inst{1}\orcidID{0000-0002-7373-4959} \and
Kristy Elizabeth Boyer\inst{1}\orcidID{0000-0003-3434-3450} \and
Maya Israel\inst{1}\orcidID{0000-0003-0302-6559}\and
Shiyan Jiang\inst{2}\orcidID{0000-0003-4781-846X}
}

\authorrunning{S. Zhang et al.}
%

\institute{
University of Florida, Gainesville, FL, United States \\
\email{\{zhangshan, a.botelho, keboyer, misrael\}@ufl.edu}
\and
University of Pennsylvania, Philadelphia, PA, United States \\
\email{\{azamb13, jiang33\}@upenn.edu}
\and
North Carolina State University, Raleigh, NC, United States \\
\email{xtian9@ncsu.edu}
\and
The University of Tennessee, Knoxville, TN, United States \\
\email{ysong51@utk.edu}
}

\maketitle              
\begin{abstract} 
As Artificial Intelligence (AI) education has become a key component of K–12 curricula, activities such as designing and developing conversational agents are increasingly used as instructional practice. Prior work has primarily examined these activities by focusing on students' learning outcomes or the quality of final AI artifacts, offering limited insight into the collaborative processes through which learning unfolds during AI system development. Although the AIED community has a long history of studying collaborative learning in STEM and Computing education, the emergence of AI learning environments in which students build AI systems presents new opportunities to understand how collaboration unfolds in AI education contexts. Grounded in these foundational works, the current study examines collaborative interaction among middle school students engaged in the design and development of an AI chatbot. Using Ordered Network Analysis of students' dialogue and development actions, we characterize how collaboration is organized over time and how interaction patterns relate to chatbot quality and AI knowledge outcomes. Results reveal that higher-quality chatbots are associated with more integrated sequences linking explanation, testing, and refinement. Interaction patterns involving articulated reasoning and repeated testing and revision in response to chatbot output were also associated with stronger AI knowledge outcomes. These findings provide a process-oriented account of collaborative AI chatbot development and extend AIED research on collaborative learning processes to AI education contexts.

\keywords{Chatbot Development \and Collaborative Learning \and AI Literacy \and Ordered Network Analysis \and AI Education.}
\end{abstract}
\section{Introduction}

Artificial intelligence (AI) has become increasingly embedded in everyday life, shaping how children and youth interact with technology \cite{song2024framework}. Encompassing the understanding of core AI concepts, the use and creation of AI systems, and engagement with the ethical and societal implications of AI, the cultivation of AI literacy has emerged as a critical goal in K--12 education \cite{long2020ai}. In response, frameworks for K--12 AI education have been developed to articulate what students should learn \cite{touretzky2023machine}, how AI can be taught through developmentally appropriate pedagogical approaches \cite{dogan2025artificial}, and why AI education matters for students' future civic participation and workforce readiness \cite{song2024framework,touretzky2019envisioning}.

Building on these frameworks, prior work has implemented a wide range of AI learning experiences across formal and informal educational settings, reporting positive relationships with students' AI conceptual understanding \cite{chiu2021creation}, skills and performance \cite{park2024implementing}, ethical awareness \cite{ali2021children}, and non-cognitive outcomes such as engagement and interest \cite{xia2022self}. A common feature of many of these experiences is the emphasis on project-based, design-oriented activities in which students collaboratively design and refine AI artifacts such as classifiers or conversational agents \cite{pearce2023build,tian2024designing}. Although this body of work has demonstrated what students can learn through collaborative AI design activities, it has offered more limited insight into how learning unfolds during the collaborative processes through which AI systems are designed and developed. 

A large body of research in the Artificial Intelligence in Education (AIED) and Computer-Supported Collaborative Learning (CSCL) communities has examined how learning unfolds through collaboration across a range of domains, including programming \cite{carpenter2020detecting,rodriguez2013repairing}, problem solving \cite{earle2023confusion,rodriguez2015discovering}, and other open-ended, design-oriented activities \cite{rodriguez2013repairing,tsan2018alright}. Central to this work is a focus on understanding learning \emph{as it unfolds} through moment-by-moment interaction, with prior studies modeling how collaborators coordinate dialogue, regulate joint activity, and construct shared understanding over time \cite{earle2023confusion,rodriguez2015discovering,rodriguez2013repairing,roll2016evolution}. This literature provides strong foundations for studying collaborative learning as a temporally organized, interactional process, and offers guidance for analyzing how interaction patterns relate to learning and task outcomes.

The collaborative design and development of AI systems differs from many other domains in that AI artifacts often exhibit probabilistic, opaque, and emergent behaviors that are difficult for novices to predict or explain. Engaging productively with such systems therefore requires learners to interpret unexpected outputs, generate and test hypotheses, and negotiate shared explanations of system behavior through interaction \cite{stahl2002contributions,stahl2010guiding}. In collaborative design settings, these sensemaking activities are distributed across dialogue and coordinated engagement with the AI system, making interaction a central locus of learning. However, much of the current AI education literature has primarily operationalized learning through post-tests or evaluations of final AI artifacts \cite{weng2024integrating,yim2025artificial,yoder2020gaining}. As a result, comparatively fewer studies have examined how students' collaborative dialogue and system-oriented development actions are jointly organized over time during AI system design and development, or how such interactional patterns relate simultaneously to both design quality and AI knowledge outcomes.

To address this gap, the present study examines collaborative interaction during middle school students' AI chatbot design and development, focusing on how patterns of dialogue and development actions unfold over time and relate to both development and learning outcomes. Specifically, we address the following research questions:
\begin{itemize}
    \item \textbf{RQ1}: How do students' collaborative dialogue and AI development actions co-occur and sequence during the chatbot design and development?
    \item \textbf{RQ2}:  How do these interactional patterns differ between groups that produce higher- and lower-quality AI chatbots?
    \item \textbf{RQ3}:  How are different patterns of collaborative interaction associated with students' AI knowledge outcomes?
\end{itemize}

\section{Related Work}

AIED and CSCL research has extensively examined how collaborative learning unfolds during open-ended tasks such as pair programming and coordinated problem-solving activities, where learners jointly plan, implement, and refine artifacts \cite{carpenter2020detecting,rodriguez2015discovering,rodriguez2013repairing}. Prior work has shown that students' dialogue plays an important role in coordinating roles, negotiating meaning, and regulating joint activity \cite{earle2023confusion}. Studies of small-group collaboration have identified dialogue acts such as suggestion, uptake, and elaboration that support coordination, alongside patterns of imbalanced participation that can hinder collaborative engagement \cite{tsan2018alright,tsan2021}. 

Beyond static dialogue categorizations, AIED research has emphasized modeling collaboration as a temporally organized process. Sequence- and state-based approaches demonstrate that collaborative activity involves transitions among interactional states such as exploratory talk, confusion, and disagreement, with reasoning often interrupted by breakdowns before returning to exploration \cite{earle2023confusion}. Modeling these temporal dynamics has allowed researchers to distinguish productive and unproductive collaboration trajectories and to identify patterns associated with learning outcomes \cite{rodriguez2015discovering,rodriguez2013repairing}. Recent work using interpretable temporal clustering further shows that sequences linking explanation, testing, and refinement are associated with stronger learning outcomes \cite{kim2025collaborative}. Although these approaches have provided powerful tools for analyzing collaborative interaction, they have often focused on a single stream of activity (e.g., dialogue), offering limited insight into how multiple forms of activity (e.g., artifact development) are coordinated over time. 

In AI education research, collaborative AI system design and development have been widely adopted in design-oriented project-based learning environments where students iteratively develop and test AI artifacts such as chatbots and classifiers \cite{kokotsaki2016project,tian2024examining}. Across these studies, learning has often been operationalized through outcome-oriented measures, including pre- and post-tests targeting AI concepts such as training data and classification \cite{chiu2021creation,park2024implementing,xia2022self}, surveys capturing attitudinal or ethical outcomes \cite{ali2021children,xia2022self}, and rubric-based evaluations of final AI artifacts \cite{ali2021children,park2024implementing}. Recent syntheses of the AI education literature similarly note a predominant reliance on post-hoc assessments and artifact evaluations to characterize learning \cite{zhang2025learning}. Although these approaches have demonstrated learning gains, they provide limited insight into the interactional mechanisms through which learning occurs during the design and development of AI systems, such as how students collaboratively reason about system behavior, interpret system feedback, or coordinate design actions over time \cite{weng2024integrating,yim2025artificial,yoder2020gaining}.

\section{Methodology}
\subsection{Participants and Study Procedure}
Data were collected from a public middle school in the southeastern United States during the Spring 2024 semester. Across six science classes ($N = 128$ students), parental consent and student assent were obtained for 100 students; all procedures were approved by the institutional IRB. Of the 97 students who reported demographic information, 49 identified as girls, 46 as boys, one as non-binary, and one preferred not to disclose. Students identified as 38 Asian, 34 White, 20 Black/African American, 6 Hispanic/Latinx, 5 self-described, 3 Native American, and 3 preferred not to disclose race/ethnicity (multiple selections allowed). The mean age was 11.7 years ($SD = 0.48$).

\begin{figure}[t]
    \centering
    \includegraphics[width=.95\linewidth]{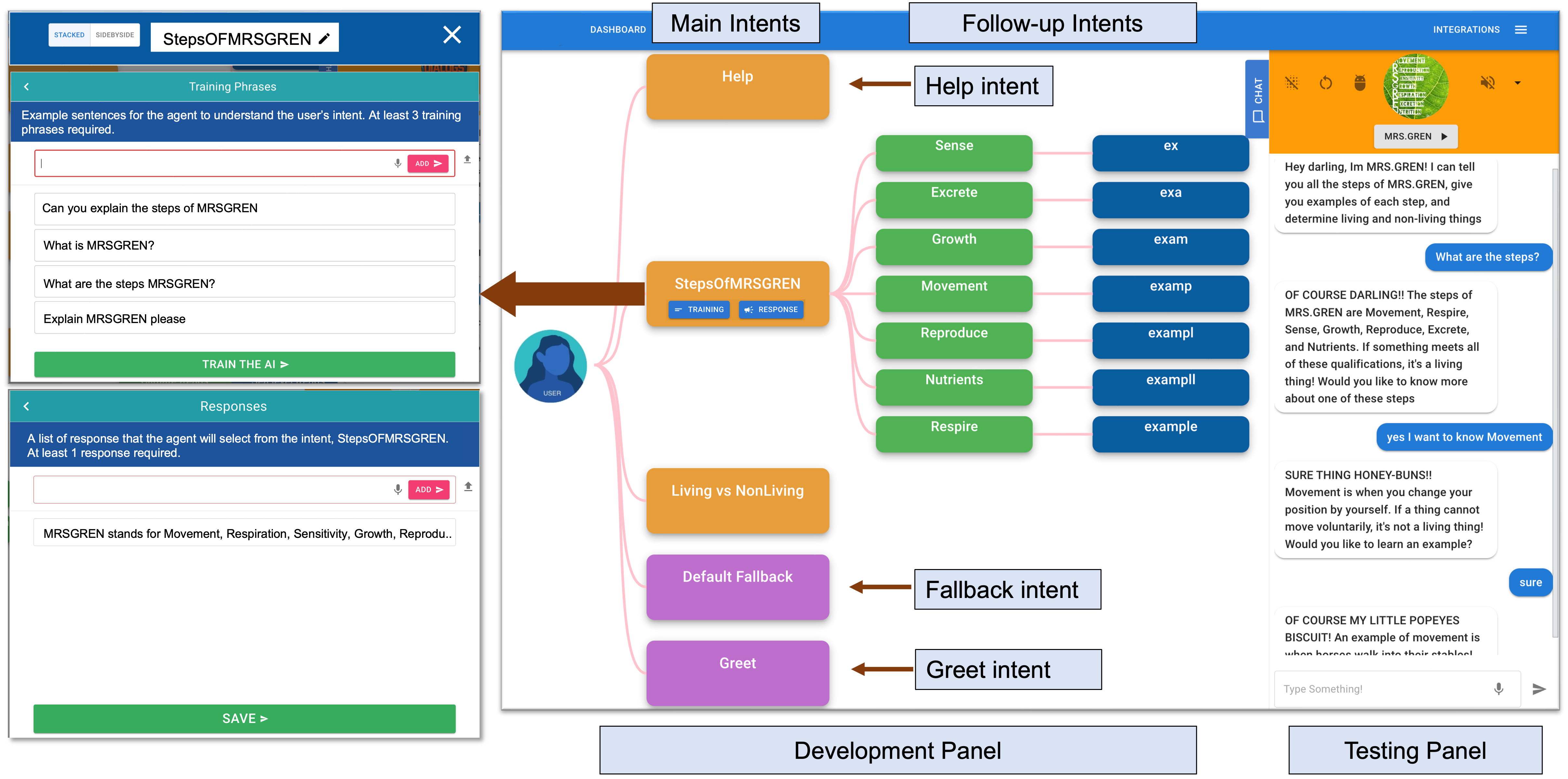}
    \caption{Learning Environment with an example chatbot}
    \label{fig:AMBY}
\end{figure}

The study spanned ten 50-minute class sessions over four weeks. During initial sessions, students completed a pre-survey reporting prior programming experience (i.e., whether they had written a program before) and received an introduction to AI, conversational AI, and the chatbot-building environment used in the study, ``AI Made By You'' (AMBY). AMBY is a web-based learning environment that supports chatbot design by enabling users to define intents (main and follow-up), add training phrases and responses for intents, and iteratively test and refine chatbot behavior \cite{tian2023amby}. Figure~\ref{fig:AMBY} shows the interface.

Following the introductory sessions, students were randomly assigned to pairs and collaborated on scientific chatbot development in AMBY using a pair-programming approach, alternating between ``driver'' and ``navigator'' roles \cite{earle2023confusion}. Student-created chatbot artifacts, including structured representations of intents, training phrases, and responses, were evaluated using a researcher-developed rubric assessing ten dimensions (e.g., intent structure, conversation design, training phrases, and response quality-scoring), with interrater reliability of $\kappa = 0.83$ across dimensions \cite{tian2024examining}. Details about the scoring rubric are available on our Open Science Framework (OSF) repository\footnote{\url{https://osf.io/ga2bk/overview?view_only=5f076ab2f04144738bc3815ee49310df}}. Rubric scores were aggregated into an overall \textbf{chatbot quality score} for each project. At the conclusion of the activity, students completed a paper-and-pencil \textbf{AI knowledge post-assessment} (available on OSF) consisting of 15 items (14 multiple-choice, one open-ended), each aligned with a specific learning objective related to AI and conversational agents.



\subsection{Dialogue Tagging}

We analyzed collaborative dialogue and chatbot artifacts from 47 student pairs participating in pair programming sessions. Audio and video recordings were captured using laptop-based recording software, and human transcriptions were obtained through a third-party service (https://www.rev.com/). The resulting corpus includes 32{,}976 utterances across 47 dyads, with an average of 701.62 utterances per dyad (SD = 293.61, min = 174, max = 1{,}514).

To examine how students collaborated while co-constructing chatbot artifacts, we adapted an established dialogue coding scheme from prior AIED research on collaborative learning in pair programming \cite{earle2023confusion}. The scheme draws on Mercer's exploratory talk framework \cite{mercer2002words} and a dialogue act taxonomy developed by Zakaria et al. \cite{zakaria2022two} to characterize collaborative discourse during programming activities. Grounded in learning sciences perspectives, the scheme captures fine-grained dialogue acts reflecting key collaborative processes, including idea generation, sense-making and reasoning, coordination and regulation, and affective or socio-relational actions. These dialogue acts characterize how learners coordinate ideas, negotiate decisions, and regulate joint activity during project-based collaboration. Because differentiating fine-grained dialogue acts required contextual information beyond transcripts alone, dialogue tagging was conducted while simultaneously viewing classroom video recordings.

To establish reliability, two researchers independently coded 20\% of the data, achieving a Cohen's kappa of 0.84, indicating strong inter-rater agreement \cite{sun2011meta}. Discrepancies were resolved through discussion, after which one researcher completed the remaining coding. Table~\ref{tab:collaborative_dialogue_scheme} presents the full set of dialogue categories used in this study; detailed definitions and examples are available on OSF.

\begin{table}[t]
\centering
\setlength{\tabcolsep}{6pt}
\renewcommand{\arraystretch}{0.95}
\caption{Collaborative dialogue coding scheme with relative frequencies ($N = 32{,}976$)}
\label{tab:collaborative_dialogue_scheme}
\begin{tabular}{p{0.3\linewidth} p{0.10\linewidth} p{0.3\linewidth} p{0.10\linewidth}}
\hline
\textbf{Dialogue} & \textbf{Freq.} & \textbf{Dialogue} & \textbf{Freq.} \\
\hline
Explanation/justification & 15.22\% & Checking & 1.62\% \\
Other & 12.72\% & Disagreement & 1.51\% \\
Directive & 12.48\% & Brainstorming & 1.40\% \\
Question--other & 9.13\% & Acknowledge & 0.97\% \\
Repeat & 7.77\% & Help-seeking & 0.89\% \\
Agreement/confirmation & 6.96\% & Confusion & 0.73\% \\
Suggestion & 6.75\% & Read aloud & 0.68\% \\
Chatbot response & 6.64\% & Voice command & 0.68\% \\
Off-task & 5.48\% & Error correction & 0.34\% \\
Facilitator guidance & 5.26\% & Antagonistic action & 0.18\% \\
Justified Disagreement & 2.35\% & Frustration & 0.09\% \\
Reference & 0.08\% & High-order question & 0.06\% \\
\hline
\end{tabular}
\end{table}

In addition to the collaborative dialogue coding scheme, we developed an artifact development coding scheme to capture students' behaviors as they designed their chatbots. This scheme records what students were doing in the AI chatbot-building process as they spoke. The same coding procedure was applied: two researchers jointly reviewed the videos and independently coded 20\% of the data, achieving a Cohen's kappa of 0.89, indicating strong agreement. After resolving discrepancies through discussion, one researcher coded the remaining data. Table~\ref{tab:chatbot_codes_freq} presents the AI chatbot development coding scheme (further explanations and examples are available on our OSF repository).

\begin{table}[t]
\centering
\setlength{\tabcolsep}{6pt}
\renewcommand{\arraystretch}{0.95}
\caption{Chatbot development codes and relative frequencies ($N=32{,}976$)}
\label{tab:chatbot_codes_freq}
\begin{tabular}{p{0.32\linewidth} p{0.1\linewidth} p{0.34\linewidth} p{0.08\linewidth}}
\hline
\textbf{Code} & \textbf{Freq.} & \textbf{Code} & \textbf{Freq.} \\
\hline
Other & 21.89\% & Debugging & 2.29\% \\
Testing AI Chatbot & 13.74\% & Add train phrases (main) & 2.18\% \\
Search online & 7.77\% & Add responses (greet) & 1.26\% \\
Dev responses (follow-up) & 7.06\% & Dev responses (fallback) & 1.25\% \\
Dev responses (main) & 5.59\% & Dev train phrases (greet) & 1.08\% \\
Add responses (follow-up) & 5.19\% & Add responses (fallback) & 0.89\% \\
Chatbot identity setup & 3.97\% & Explore interface & 0.85\% \\
Dev train phrases (follow-up) & 3.92\% & Develop entity & 0.75\% \\
Add responses (main) & 3.41\% & Understanding AI logic & 0.50\% \\
Setup follow-up intent & 3.29\% & Adding entity & 0.49\% \\
Dev train phrases (main) & 2.94\% & Dev train phrases (fallback) & 0.46\% \\
Add train phrases (follow-up) & 2.92\% & Add train phrases (greet) & 0.33\% \\
Setup main intent & 2.45\% & Implement entity & 0.25\% \\
Dev responses (greet) & 2.44\% & Setup entity & 0.25\% \\
Greet intent & 0.02\% & Explore entity & 0.18\% \\
Add responses (help) & 0.02\% & Issue resolved & 0.15\% \\
Dev responses (help) & 0.02\% & Info gathering & 0.05\% \\
Add train phrases (help) & 0.05\% & Add train phrases (fallback) & 0.05\% \\
Dev train phrases (help) & 0.04\% &  &  \\
\hline
\end{tabular}
\end{table}

\subsection{Data Analysis} 
To examine differences in interaction patterns associated with chatbot performance, we conducted an Ordered Network Analysis (ONA; \cite{tan2022ordered}) on both collaborative dialogue acts and AI chatbot development behaviors. ONA uses a moving window to identify connections among constructs—dialogue acts in the collaboration dialogue coding scheme and student behaviors in the AI chatbot coding scheme—based on their co-occurrence within the window, while explicitly accounting for their sequential order. Specifically, ONA distinguishes between the strength of a connection when dialogue act A is followed by dialogue act B (e.g., the group provides an explanation or justification and then explicitly states what to do next) versus when dialogue act B is followed by dialogue act A (e.g., the group explicitly states what to do next before providing an explanation). ONA also captures self-transitions, which reflect repeated occurrences of the same construct (e.g., providing multiple explanations consecutively and recurrently).

After identifying these ordered co-occurrences, ONA normalizes the transition counts using cosine normalization and constructs a network that represents the strength of connections among constructs (dialogue acts or behaviors) for each unit of analysis (e.g., individual students or collaboration pairs). In this network, nodes represent constructs, node size reflects the frequency of self-transitions (i.e., construct repetition), and edges represent the strength of connections between different constructs. Both self-transition strengths and transitions between different constructs are quantified on the same scale, and their values are referred to as connection weights (CW).

Each unit's network is embedded in a two-dimensional space, where the relative proximity of nodes and the centroid of each unit's network indicate which constructs and connections are more frequent among particular students or groups. These networks can also be aggregated into higher-level groups (e.g., high-performing versus low-performing students), enabling statistical comparisons between groups. In addition, the connection weights for each unit can be used in further statistical analyses alongside external measures.

The ONA models were created using WebENA \cite{marquart2018epistemic}. Groups were divided according to their chatbot performance scores using a median split. Groups with scores at or below the median (3.19) were categorized as the low chatbot performance group ($N = 25$), whereas groups with scores above the median were categorized as the high chatbot performance group ($N = 22$). For both coding schemes (collaborative dialogue acts and AI chatbot development behaviors), we used a moving window of size four to identify code co-occurrences. This window size is commonly used in similar analysis \cite{siebert2017search} and was further validated through qualitative inspection of our data. We also tested multiple window sizes ranging from 2 to 10 without observing substantive changes in the overall patterns or group differences. Each group's data were segmented by day, reflecting the assumption that discourses or behaviors occurring on the same day are more strongly connected than those separated by multiple days. Finally, for visualization purposes, we excluded codes that were infrequent in the data and did not exhibit meaningful differences between the groups under study (CW < 0.01).

To examine whether chatbot performance was associated with students' prior programming experience, we analyzed the distribution of students' responses to the pre-survey item, whether students had prior experience with written programming, across the high and low chatbot score groups. Responses to this item were cross-tabulated with chatbot performance group (high vs.\ low) to assess potential differences in prior programming experience between groups. This analysis allowed us to evaluate whether observed differences in chatbot performance were potentially related to students' prior programming exposure rather than differences in collaborative interaction processes.

We examined associations between students' AI knowledge and interaction patterns using Spearman's rank-order correlations. Because the AI knowledge assessment was designed around the chatbot system and students had no prior experience with this system, individual scores were aggregated at the group level ($M = 25.8$, $SD = 5.71$) and correlated with ONA-derived connection weights capturing the strength of ordered transitions among collaborative dialogue acts and AI chatbot development behaviors. Spearman's correlation was used to accommodate non-normally distributed network weights and assess monotonic relationships between AI knowledge and interaction dynamics. To control for false discoveries, we applied a Benjamini–Hochberg correction \cite{benjamini1995controlling} to individual correlations and conducted a Monte Carlo analysis \cite{metropolis1949monte} with 10{,}000 iterations to evaluate whether the overall pattern of observed correlations was unlikely to have occurred by chance.

\section{Results}

\begin{figure}[!b]
    \centering
    \includegraphics[width=0.85\linewidth]{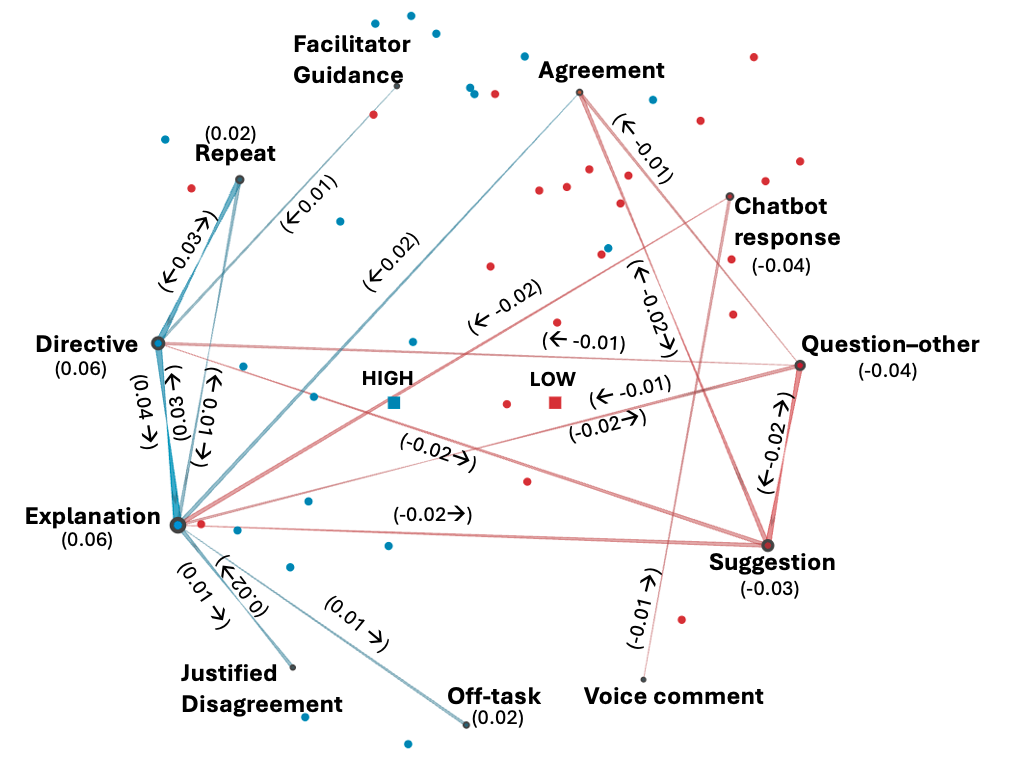}
    \caption{Ordered Network Analysis comparing collaborative dialogue acts between groups with high (blue) and low (red) chatbot performance.}
    \label{fig:col_ona}
\end{figure}

Figure~\ref{fig:col_ona} presents the ordered network analysis (ONA) comparing collaborative dialogue patterns between groups with high and low chatbot performance (CWs for both coding schemes are available on our OSF repository). Across both groups, interaction networks are primarily organized around the dialogue acts of \textit{explanation} and \textit{directive}, which exhibit the strongest self-loop connection weights (CW $>$ 0.2). These dialogue acts capture students' articulation of reasoning and clarification of ideas, as well as utterances used to guide or coordinate partner actions during collaborative AI chatbot construction. A Mann--Whitney U test comparing the two ordered network models reveals significant differences in the dialogue structure between the groups ($p < 0.001$).

As shown in Figure~\ref{fig:col_ona}, groups with high chatbot scores exhibit higher connection weights for \textit{explanation} (CW = 0.345 vs.\ 0.282), \textit{directive} (CW = 0.257 vs.\ 0.201), \textit{repeat} (CW = 0.119 vs.\ 0.100), and \textit{off-task} (CW = 0.249 vs.\ 0.225), along with stronger ordered transitions among these dialogue acts. Within the coding scheme, \textit{repeat} reflects the verbal repetition of typed input or previously stated content to confirm or reinforce actions, while \textit{off-task} captures brief departures from the task that do not directly advance chatbot development. Bidirectional transitions between \textit{directive} and \textit{explanation} are more pronounced in high-performing groups (directive $\rightarrow$ explanation: CW = 0.175 vs.\ 0.139; explanation $\rightarrow$ directive: CW = 0.181 vs.\ 0.152), indicating recurrent coordination between action guidance and articulated reasoning during chatbot construction.

Groups with high chatbot scores exhibited stronger transitions involving \textit{justified disagreement}, including self-loops (CW = 0.051 vs.\ 0.034) and transitions to \textit{explanation} (CW = 0.051 vs.\ 0.034), reflecting disagreement accompanied by articulated reasoning. By contrast, groups with low chatbot scores showed higher connection weights for \textit{question--other} (CW = 0.134 vs.\ 0.093), \textit{chatbot response} (CW = 0.181 vs.\ 0.146), and \textit{suggestion} (CW = 0.084 vs.\ 0.050), indicating greater emphasis on procedural questioning, proposal generation, and orientation toward system output. Overall, high-performing groups demonstrated more densely connected transitions among explanation, directive, repetition, and disagreement acts, whereas low-performing groups exhibited interaction patterns centered on procedural and chatbot-response-focused sequences.


\begin{figure}[ht]
    \centering
    \includegraphics[width=0.85\linewidth]{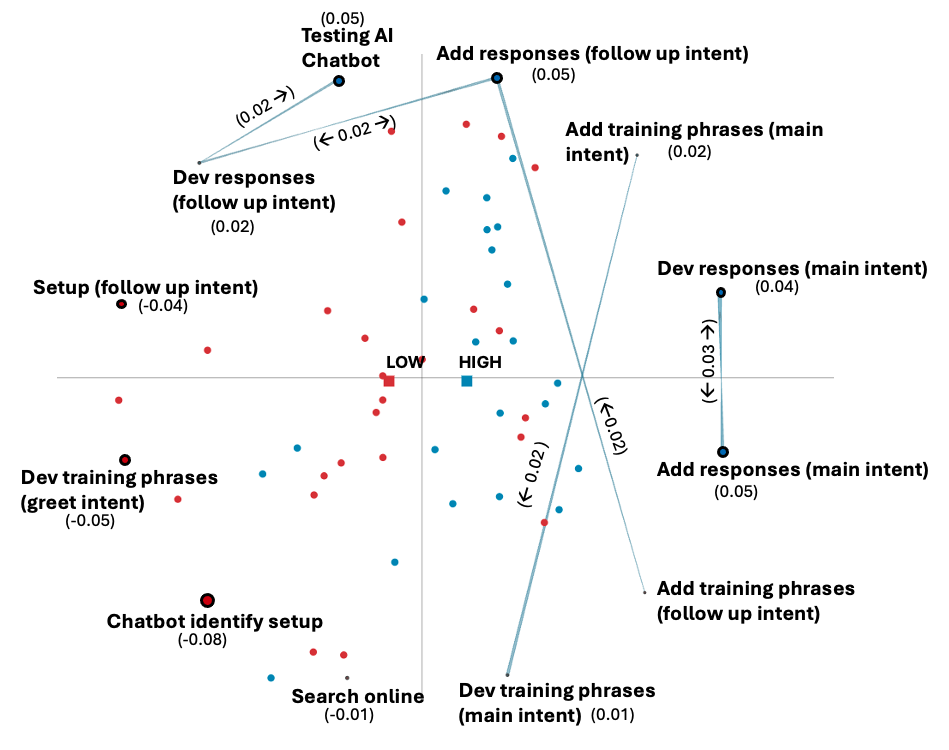}
    \caption{Ordered Network Analysis comparing AI chatbot development behaviors between groups with high (blue) and low (red) chatbot performance.}
    \label{fig:bot_ona}
\end{figure}

Figure~\ref{fig:bot_ona} presents the ordered network comparison of AI chatbot development behaviors between groups with high and low chatbot performance. In both groups, the strongest self-loop connections occur for \textit{testing} and \textit{developing responses} (CW $>$ 0.15), reflecting repeated engagement in evaluating chatbot output and refining responses. A Mann--Whitney U test indicates significant differences in the structure of development behaviors between the groups ($p = 0.015$).

Groups producing higher-quality chatbots exhibited stronger self-loops for \textit{testing} (CW = 0.564 vs.\ 0.511), \textit{developing responses for the main intent} (CW = 0.229 vs.\ 0.187), and \textit{adding responses for both main and follow-up intents} (CW = 0.121 vs.\ 0.074; CW = 0.180 vs.\ 0.134), alongside more frequent transitions among development and testing activities. These patterns reflect iterative movement across response conceptualization, implementation, and evaluation. In contrast, groups with lower chatbot scores showed higher self-loop weights for activities such as \textit{chatbot identity setup} (CW = 0.207 vs.\ 0.123), \textit{developing training phrases for the greet intent} (CW = 0.077 vs.\ 0.028), and \textit{setting up follow-up intents} (CW = 0.124 vs.\ 0.083), indicating extended engagement within individual development stages and fewer transitions into testing and response refinement. 

Importantly, differences in chatbot performance were not explained by prior programming experience. Although the low-performing group included more students reporting prior programming experience on both survey items (\textit{Yes, Yes}: $n = 4$), the high-performing group included students with mixed, partial, or no prior experience (e.g., \textit{Yes, No}: $n = 6$; \textit{Yes, Don't Know}: $n = 3$; \textit{No, No}: $n = 3$), suggesting that observed performance differences are more closely related to interactional patterns than to prior experience alone. 

Consistent with this interpretation, associations between interaction patterns and AI knowledge outcomes revealed a non-random structure. Across 250 Spearman correlations examining collaborative dialogue and development behaviors, 31 yielded p-values below 0.05 (all results available on the OSF repository). Although none remained significant after Benjamini--Hochberg correction, the Monte Carlo analysis indicated that observing this number of correlations by chance is unlikely ($p $ < $ 0.001$; 95\% CI: 0 to 20 significant results due to chance), suggesting that interaction patterns during collaborative AI chatbot development are meaningfully related to students' AI knowledge outcomes.

For collaborative dialogue, AI knowledge outcomes were positively associated with interaction sequences centered on \textit{explanation} and \textit{repeat}, including bidirectional connections between \textit{explanation} and \textit{repeat} ($\rho = 0.417$; $\rho = 0.337$) and between \textit{explanation} and \textit{chatbot response} ($\rho = 0.357$; $\rho = 0.343$). In contrast, AI knowledge outcomes were negatively associated with sequences dominated by \textit{directive} and \textit{suggestion} acts, including \textit{directive}--\textit{suggestion} ($\rho = -0.467$; $\rho = -0.418$), \textit{suggestion} self-loops ($\rho = -0.333$), \textit{suggestion}--\textit{question--other} ($\rho = -0.319$; $\rho = -0.314$), as well as \textit{directive} self-loops ($\rho = -0.288$) and \textit{directive}--\textit{question--other} transitions ($\rho = -0.291$). Thus, these patterns indicate that collaborative dialogue supporting explanation, repetition, and mutual grounding is more strongly associated with AI knowledge gains, whereas interaction sequences emphasizing procedural direction may limit opportunities for conceptual understanding.

For chatbot development behaviors, higher AI knowledge outcomes were associated with stronger \textit{testing} self-loops ($\rho = 0.372$) and refinement-oriented transitions, including \textit{develop responses for main intent}--\textit{testing} ($\rho = 0.308$) and \textit{testing}--\textit{search online} ($\rho = 0.293$). In contrast, several of the strongest negative associations involved connections with \textit{adding training phrases for follow-up intent}, including its coupling with \textit{adding responses for follow-up intent} ($\rho = -0.496$), \textit{develop responses for follow-up intent} ($\rho = -0.454$), \textit{setup follow-up intent} ($\rho = -0.377$), and \textit{adding responses for main intent} ($\rho = -0.344$), characterizing how development activity patterns align with AI knowledge outcomes. Overall, iterative testing and refinement of core intents align with stronger AI knowledge outcomes, whereas extensive focus on follow-up intent configuration may be less conceptually productive.

\section{Discussion} 

This study investigated how collaborative interaction unfolds during AI chatbot design and development and how these interactional patterns relate to design and learning outcomes. Our findings characterize collaborative AI chatbot design and development as a temporally organized process in which dialogue and development actions are closely intertwined (RQ1). Across groups, students' activity was structured around recurring interactional sequences that linked explanation, directive coordination, and engagement with chatbot behavior. These patterns indicate that collaborative AI chatbot design and development involve ongoing cycles of articulating ideas, testing responses, and revising designs, rather than a more linear progression. This aligns with AIED work showing that collaborative learning can be productively represented as transitions among problem-solving modes over time, rather than as isolated frequencies of talk moves \cite{earle2023confusion,rodriguez2015discovering}.

Examining differences between groups (RQ2), the analyses reveal that groups whose chatbots demonstrated higher quality exhibited more interconnected dialogue and development patterns, with frequent transitions among explanation, testing, and response refinement. In contrast, lower-quality chatbots were associated with interaction patterns that emphasized more segmented or procedural trajectories of activity. These differences were not explained by students' prior programming experience, suggesting that how students coordinated their interaction during design played an important role in shaping design outcomes.

By analyzing students' AI knowledge outcomes (RQ3), we observe that interaction sequences involving articulated reasoning, repetition of prior contributions, and engagement with chatbot responses were positively associated with AI knowledge scores, whereas patterns dominated by directive or suggestion-oriented exchanges showed negative associations. Although these relationships do not establish causality, they point to alignment between how collaborative interaction is organized and variation in students' learning outcomes.

\section{Limitations and Future Work}

Several limitations should be considered when interpreting these findings. First, the analyses focus on middle school students engaged in a single AI chatbot design and development, which may limit generalizability to other age groups, learning contexts, or AI tools, where collaborative interaction may take different forms. Second, although the analytic approach captures fine-grained temporal coordination between dialogue and development actions, it does not account for all aspects of interaction (e.g., gesture, affect). Finally, the observed associations remain correlational and therefore do not establish causal relationships. Future work could extend this approach by incorporating additional data modalities, examining longitudinal trajectories across activities, exploring how instructional supports impact interactional organization, and further developing scalable and interpretable analytic methods for studying collaborative AI learning processes.

\section{Conclusion}

This study examined collaborative interaction during middle school students' AI chatbot design and development by analyzing the temporal coordination of students' dialogue and development actions. Interpreted through a collaborative learning perspective, the findings highlight the role of explanation, coordination, and iterative engagement with system behavior in shaping both design and learning outcomes. From a design perspective, supporting interactional cycles that integrate explanation, testing, and revision may foster more coordinated collaborative AI system design and development. From a research perspective, the results underscore the value of integrating multiple process data streams to better capture how learning unfolds during AI system design and development. Overall, this work builds on previous research on collaboration by modeling the temporal coordination of dialogue and development actions, extending process-oriented analyses to the context of collaborative AI system design and development.

\begin{credits}
\subsubsection{\ackname}
This material is based upon work supported by the National Science Foundation and the Institute of Education Sciences under Grants DRL-2229612, DRL-2048480, R305B230007, and \#2331379, as well as the Gates Foundation (\#078981), support from the Learning Engineering Tools Competition, and other anonymous philanthropy. Any opinions, findings, and conclusions or recommendations expressed in this material are those of the author(s) and do not necessarily reflect the views of the National Science Foundation or the U.S. Department of Education.
\end{credits}



%
%
%
\bibliographystyle{splncs04}
\bibliography{mybibliography}
\end{document}